%% file: Principe_HBC_accepted_twocol.tex
\documentclass{aa}  

\usepackage{graphicx}
\usepackage{txfonts}
\begin{document}

   \title{The Multiple Young Stellar Objects of HBC 515: An X-ray and Millimeter-wave Imaging Study in (Pre-main Sequence) Diversity}

%   \subtitle{test}

   \author{D. A. Principe
          \inst{1,2},
          G. G. Sacco\inst{3},
          J. H. Kastner\inst{4},
          D. Wilner\inst{5},
          B. Stelzer \inst{6},
          \and
          G. Micela\inst{6}%\fnmsep\thanks{Just to show the usage
         % of the elements in the author field}
          }

   \institute{N\'{u}cleo de Astronom\'{i}a de la Facultad de Ingenier\'{i}a, Universidad Diego Portales, Santiago, Chile              \\
              \email{daveprincipe1@gmail.com}
         \and
             Millennium Nucleus Protoplanetary Disks, Universidad Diego Portales, Chile \vspace{-3.5mm}\\
%             \email{c.ptolemy@hipparch.uheaven.space}
         \and
         INAF-Osservatorio Astrofisico di Arcetri, Largo E. Fermi, 5, 50125, Firenze, Italy \\
         \email{gsacco@arcetri.inaf.it}    
         \and
         Chester F. Carlson Center for Imaging Science, School of Physics \& Astronomy, and Laboratory for Multiwavelength Astrophysics, Rochester Institute of Technology, 54 Lomb Memorial Drive, Rochester NY 14623 USA
 \\
         \email{jhk@cis.rit.edu}    
         \and
                  Harvard-Smithsonian Center for Astrophysics, 60 Garden Street, Cambridge, MA 02138 USA \vspace{-3.5mm}\\
         \and
         INAF - Osservatorio Astronomico di Palermo, Piazza del Parlamento 1, 90134 Palermo, Italy \\
         %\email{stelzer@astropa.unipa.it; giusi@astropa.unipa.it}       
                       }

   \date{Received 18 August 2016 / Accepted 6 October 2016}

% \abstract{}{}{}{}{} 
% 5 {} token are mandatory
 
  \abstract
   {
   We present Chandra X-ray Observatory and Submillimeter Array (SMA) imaging of \object{HBC 515}, a system consisting of multiple young stellar objects (YSOs). The five members of HBC 515 represent a remarkably diverse array of YSOs, ranging from the low-mass Class I/II protostar \object{HBC 515B}, through Class II and transition disk objects (\object{HBC 515D} and C, respectively), to the ``diskless'', intermediate-mass, pre-main sequence binary HBC 515A. Our Chandra/ACIS imaging establishes that all five components are X-ray sources, with \object{HBC 515A} --- a subarcsecond-separation binary that is partially resolved by Chandra --- being the dominant X-ray source.  We detect an X-ray flare associated with HBC 515B. In the SMA imaging, HBC 515B is detected as a strong 1.3 mm continuum emission source; a second, weaker mm continuum source is coincident with the position of the transition disk object \object{HBC 515C}. These results strongly support the protostellar nature of HBC 515B, and firmly establish HBC 515A as a member of the rare class of relatively massive, X-ray luminous ``weak-lined T Tauri stars'' that are binaries and have shed their disks at very early stages of pre-MS evolution. The coexistence of two such disparate objects within a single, presumably coeval multiple YSO system highlights the influence of pre-MS star mass, binarity, and X-ray luminosity in regulating the lifetimes of circumstellar, planet-forming disks and the timescales of star-disk interactions.}


\titlerunning{The Multiple Young Stellar Objects of HBC 515}
\authorrunning{Principe et al.}
   \maketitle
%   \titlerunning{test2}
%
%-------------------------------------------------------------------

\section{Introduction}

 Young stellar objects (YSOs) are surrounded by rotating disks of gas and dust that are the likely sites of future planetary systems. The disk dissipation timescale for circumstellar disks can range from a few $\times10^5$ yr to $\sim$$10^7$ yr, and both the timescales and physical mechanisms governing this dissipation process are not well understood \citep[e.g.,][]{Williams2011}. It is apparent that the longevities of protoplanetary disks are governed by a variety of competing yet interrelated mechanisms, including viscous accretion, stellar-irradiation-driven photoevaporation, the evolution of disk gas and dust, and disk-protoplanet interactions \citep[see, e.g.,][and refs.\ therein]{Gorti2015,Andrews2015,Kastner2016}. Multiwavelength observations of young, multiple systems composed of sources formed in the same environment and (presumably) at approximately the same time are necessary to understand the importance and efficiency of these various proposed disk dissipation mechanisms. 

\begin{figure*}

\centering
\includegraphics[scale=0.44]{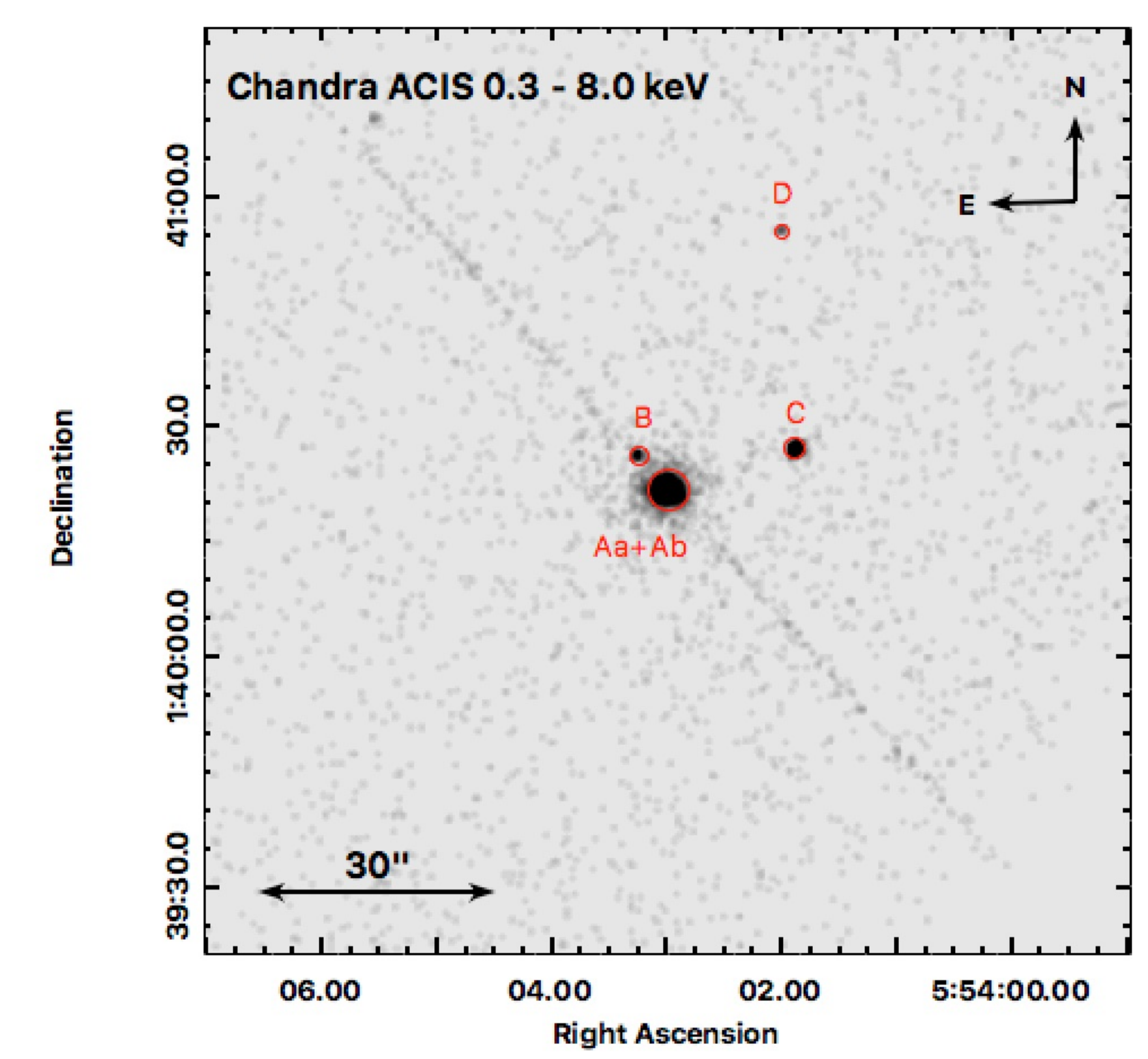}
\includegraphics[scale=0.44]{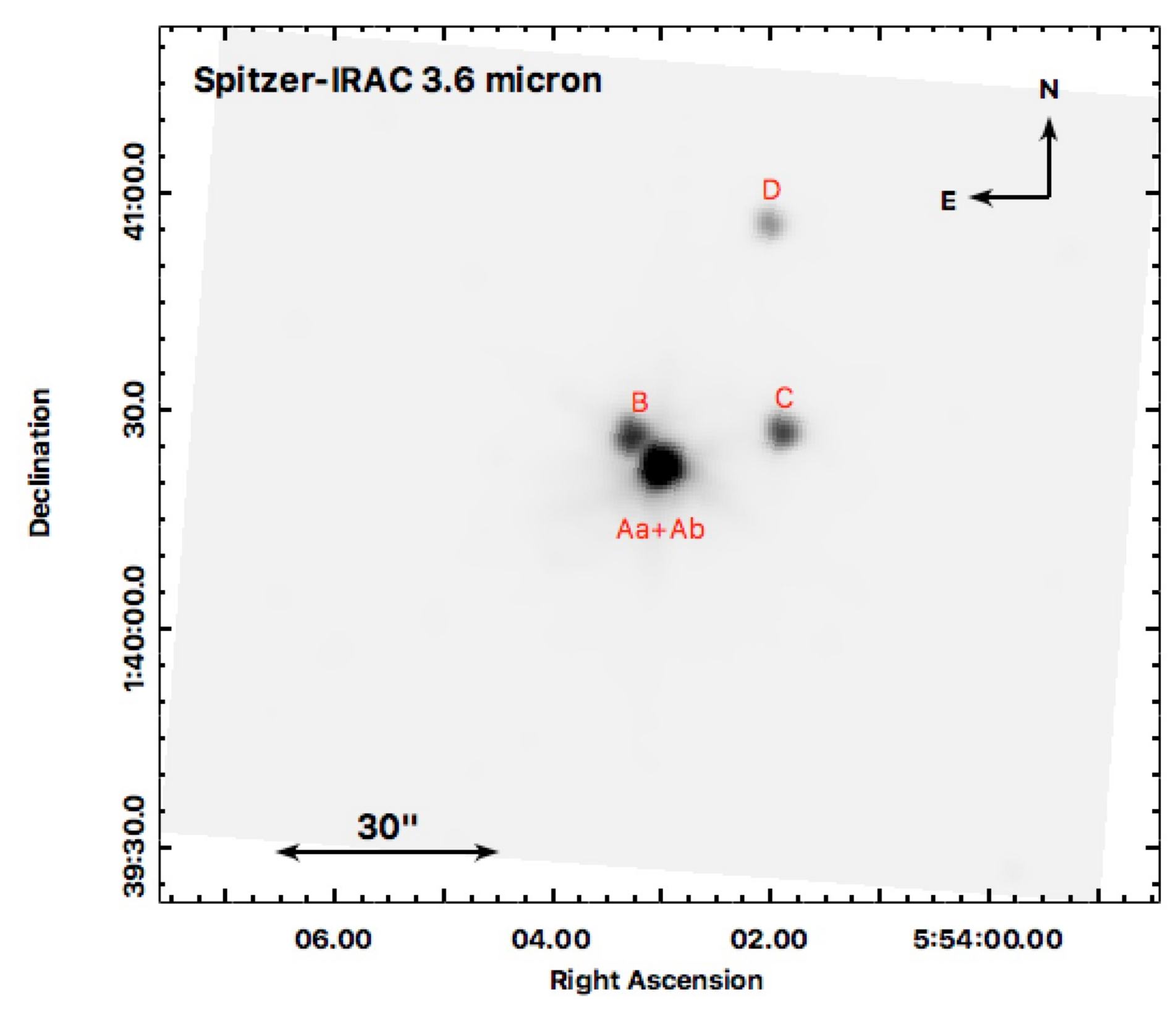}
\caption{ Comparison of Chandra ACIS-S3 X-ray (left) and Spitzer-IRAC IR (right: AOR 16214528) images of the HBC 515 system.  Individual members are indicated with red circles and annotations. The Chandra image is integrated over the energy range 0.3-8.0 keV. Coordinates are J2000. }
\label{FOV}
\end{figure*}

The multiple system HBC 515  \citep[][]{Herbig1988,Reipurth2010}, located in the \object{Lynds 1622} dark cloud at a distance of $\sim$400 pc \citep{Kun2008}, is composed of at least five objects, ranging in evolution from the embedded protostar HBC 515B to the luminous, ``diskless'', pre-MS binary HBC 515A (= \object{HD 288313}; Figure \ref{FOV}). Due to the cool photospheres of its two binary components (HBC 515 Aa and Ab), and the relative absence of strong optical emission lines and IR excess that characterize actively accreting pre-main sequence stars, \citet{Reipurth2010} classified component A as a weak-lined T Tauri star system. The two components were spatially resolved with Subaru Adaptive Optics (AO) $K$ band photometry and were found to have an angular separation of $\sim$ 0.5$''$ and a flux ratio of 1.03; combined with low-resolution spectroscopy, this flux ratio indicates both members have spectral types of $\sim$K2 \citep{Reipurth2010}.  A wavelength-dependent optical polarization signature and possible excess at  24 $\mu$m hints at the existence of a remnant circumstellar disk or envelope structure around HBC 515 Aa and/or Ab \citep{Mekkaden2007,Reipurth2010}.  From its spectral energy distribution (SED), \citet{Reipurth2010} derived an effective temperature $T_{eff} =5000$ K and luminosity $L_{bol} =22$ $L_\odot$ for HBC 515A. Assuming that this luminosity is shared equally by the two components, HBC 515A appears to be composed of two $\sim$1.8 $M_\odot$ stars with an age of only $\sim$$5\times10^5$ yr.  \citet{Reipurth2010} furthermore note the slightly elongated point spread function of component Ab in their near-IR AO imaging which, combined with the large luminosities of the components, suggests HBC 515A may be in fact be a hierarchical multiple system. 

The conclusion of a very young ($<1$ Myr) age for HBC 515A is supported by the presence of its heavily absorbed companion HBC 515B, which is separated from A by 5.4$''$. At optical/near-IR wavelengths, component B first emerges in the K band, and its brightness rapidly increases towards longer wavelengths until 24 $\mu$m, where it becomes the dominant IR source in the HBC 515 system (Figure \ref{FOV}). The fact that it emerges in the K band suggests it may be a deeply embedded (i.e., Class I) protostar \citep[][]{Reipurth2010}, although its location in a 3.6-24 $\mu$m Spitzer color-color diagram coincides with those of more evolved (i.e., Class II) young stellar objects \citep{Megeath2012}.

 \citet{Reipurth2010} suggested that another two stars, HBC 515 C and HBC 515 D, separated from the main component by 17$''$ and 36$''$, respectively (Figure \ref{FOV}), may be dynamically associated with HBC 515. Based on optical spectroscopy and their SEDs, HBC 515 C is an M4 spectral type pre-MS star likely surrounded by a transition disk (i.e., an optically thick disk with an inner gap), while HBC 515 D is a Class II YSO of spectral type M3 \citep{Kun2008,Reipurth2010,Megeath2012}. 

Here, we present Chandra X-ray Observatory and Submillimeter Array (SMA) imaging of HBC 515 to further elucidate the evolutionary state of each member and to investigate the relationship between pre-MS X-ray emission and the early evolution of circumstellar disks around such a diverse contingent of coeval YSOs. The Chandra X-ray and SMA mm-wave interferometric observations are described in \S 2, and results are presented in \S 3. We discuss the implications of these results for the natures of components A and B in \S 4, and in \S 5 we consider how this investigation, and further study of the HBC 515 system, may help inform models of the dissipation of protoplanetary disks.

%--------------------------------------------------------------------

\section{Observations and Data Reduction}

\subsection{Chandra X-ray Observatory}

A Chandra X-ray Observatory observation of the HB 515 field (Obs ID 12383, PI - G. Sacco), with an exposure time 29 ks, was obtained in January 2011 with ACIS-S in subarray mode.  A quarter-chip subarray with frame time of 0.9 seconds was chosen to alleviate the effects of pileup due to the expected high count rate of HBC 515A. The data were reduced with the Chandra Interactive Analysis software (CIAO; v.\ 4.8). The observation was energy filtered (0.3 - 8.0 keV) and time filtered using good time intervals to reduce any flaring particle background.  The CIAO tools \verb+srcflux+, \verb+dmextract+, \verb+dmstat+, \verb+glvary+ and \verb+specextract+ were used to determine background subtracted count rates, extract light curves, test for variability and perform spectral extraction on members of the HBC 515 system. Details of the spectral extraction are described in \S 3.1.

\subsection{Submillimeter Array}

Observations of the HBC 515 system at 1.3 millimeters were obtained with the Submillimeter Array (SMA) on 2012 November 13 in the compact configuration with 8 operational antennas. The weather was acceptable, with  225 GHz atmospheric opacity about 0.3 through the track, measured at the nearby Caltech Submillimeter Observatory. The digital correlator was configured to provide 4 GHz of bandwidth in each of two sidebands, with high resolution ``chunks'' of 256 channels with 0.40625 MHz spacing devoted to the $^{12}$CO and $^{13}$CO, J=2-1 lines. Complex gain calibration was enabled by a 15 minute observation loop that interleaved 10 minutes on the target with 2.5 minutes each on the nearby quasars J0532+075 and J0530+135.  Passband calibration was performed with observations of the bright quasars BL Lac and 3C279.  The absolute flux scale determined using an observation of Uranus, with an estimated accuracy of 20\%. The synthesized beam size is $\sim$ 3 $\times$ 3$''$ and the field of view is defined by the $55''$ FWHM primary beam size of the 6 meter SMA antennas at 1.3 millimeters.  Spectral data cubes were processed using MIRIAD (Multichannel Image Reconstruction Image Analysis and Display) and integrated intensity (moment 0) and velocity (moment 1) maps were created with CASA (Common Astronomy Software Applications; version 4.5).  The integrated velocity ranges for the $^{12}$CO and $^{13}$CO moment 0 maps are 9.0 and 8.3 km s$^{-1}$, respectively.    Two-dimensional elliptical gaussian fits were performed using the CASA routine \verb+imfit+ with aperture radii of $\sim$3 $''$ in order to extract source fluxes.   

\section{Results}

\subsection{Chandra X-ray images, spectra, and light curves}

In Figure \ref{FOV}, we display the 0.3--8.0 keV Chandra/ACIS-S3 X-ray image of HBC 515, alongside a Spitzer infrared 3.6 $\mu$m image of the system. As this comparison makes clear, the Chandra imaging establishes that the A, B, C, and D components of HBC 515 are X-ray sources. In Table \ref{counts_table} we list the positions and ACIS-S3 count rates for these four sources (five, including both components of HBC 515A; Figure \ref{515A_chan}), as determined within source spectral extraction regions of radii 1$''$ (corresponding to 90\% encircled energy) for components HBC 515B, C, and D and within ``core'' and ``wing'' regions in the case of the (marginally resolved) components HBC 515Aa and Ab (see next). 

\subsubsection{Decomposing Component A}

\begin{figure*}
\centering
\includegraphics[scale=0.65]{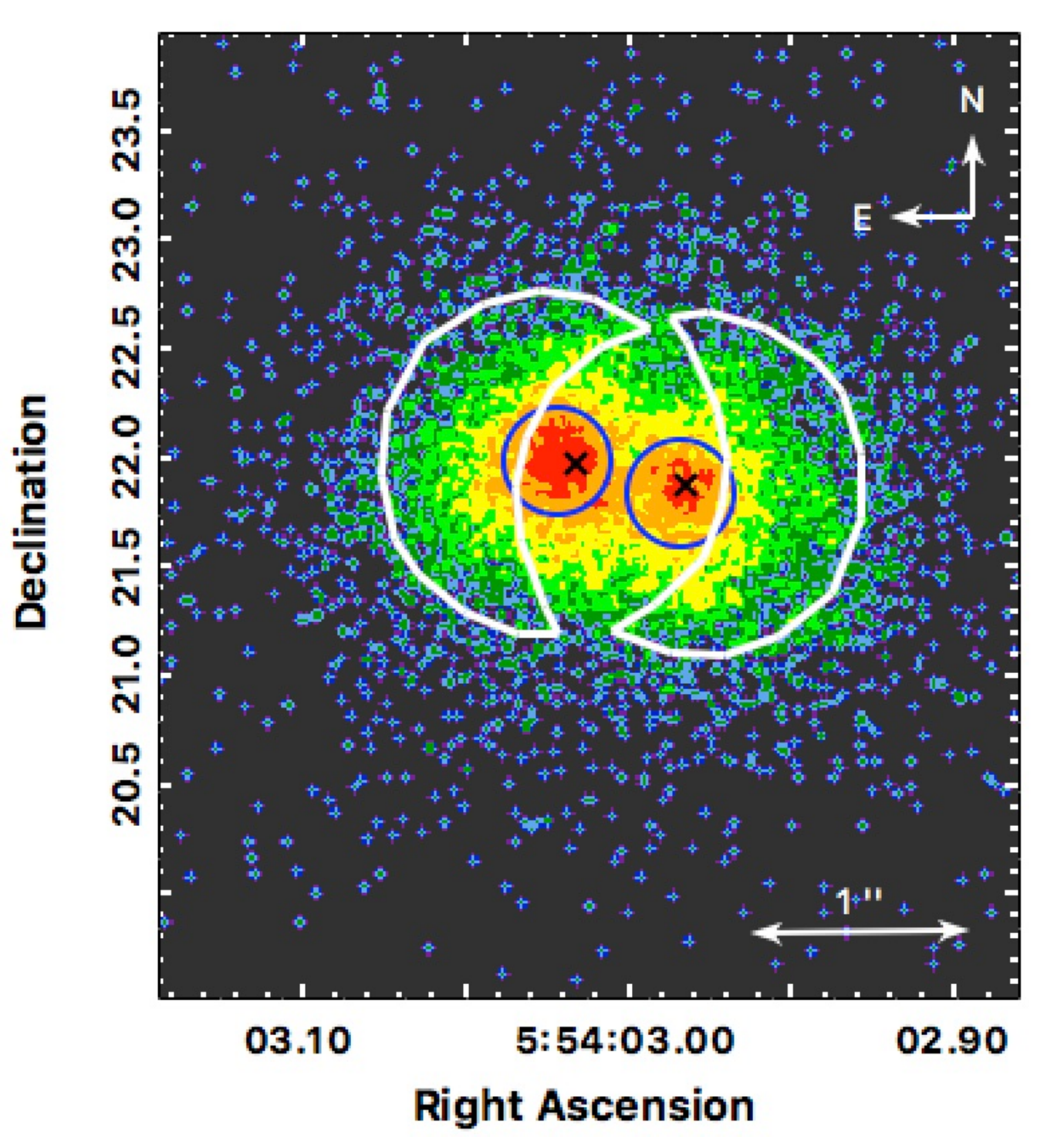}
\caption{Chandra sub-pixel view (1 sub-pixel = 0.015$\times$0.015$''$) of the HBC 515A binary with the positions of the Aa and Ab components identified with Subaru K-band AO imaging (Reipurth et al. 2010; denoted with the black X's). The location of the core and wing extraction regions are indicated as blue circles and white crescents, respectively.}
\label{515A_chan}
\end{figure*}

As is apparent both from Figure \ref{FOV} and Table \ref{counts_table}, HBC 515A (i.e., Aa and Ab combined) is by far the most X-ray luminous of the members of HBC 515. The count rate of HBC 515A is large enough to generate a CCD readout streak that partially overlays the X-ray source coincident with HBC515 B.    In Figure \ref{515A_chan}, we present a zoomed-in view of the Chandra/ACIS-S3 image of HBC 515A. The image resolution has been optimized via subpixel CCD event repositioning \citep[SER;][]{Li2003}, which is implemented as part of standard ACIS event pipeline processing. The HBC 515A binary (Aa and Ab) is partially resolved in X-rays in the SER-processed ACIS-S3 image.   The point spread functions of the two binary components overlap significantly, however, complicating their photometric and spectral decomposition. Hence, we used two spectral extraction regions (i.e., ``core'' and ``wing'') for each of the component sources, HBC 515 Aa and Ab; these extraction regions are illustrated in Figure \ref{515A_chan}. The core regions for HBC 515Aa and Ab have extraction radii of 0.25 arcseconds, corresponding to an encircled energy fraction of $\sim$40\% assuming a photon energy of $\sim$1.5 keV.  The wing extraction regions were chosen such that the location of each was at least 0.8$''$ in angular distance away from its binary counterpart, a distance where the Chandra point spread function encircled energy fraction (e.e.f.) is $\sim$90\% for sources that peak near $\sim$1.5 keV. Thus, each wing extraction region should contain $< 10\%$ of the photons from the opposite binary component, whereas each core region is likely significantly contaminated by the other component's photons.   

While the two binary components have very similar spectral types and a K-band flux ratio of 1.03 \citep{Reipurth2010}, there is {\it a priori} no reason to expect their X-ray fluxes to be so similar \citep[see, e.g.,][]{Kastner2004a,Huenemoerder2007}. Indeed, as can be seen in Figure \ref{515A_chan}, the core region of HBC 515Aa appears brighter than that of HBC 515Ab. However, given their proximity and large count rates, the core regions may suffer from both photon pileup (i.e., merging of charge from multiple photons arriving in the same or adjacent pixels within a given CCD  frame) and contamination from  the adjacent source. To investigate these effects, we performed source extraction and X-ray spectral fitting (see section 3.1.2) to compare the count rates and flux ratios of the core and wing regions of the X-ray sources associated with HBC515 Aa and Ab. We find count rate ratios of $C_{Aa}/C_{Ab}{\mathrm (core)} = 1.51$ and $C_{Aa}/C_{Ab}{\mathrm (wing)} = 1.24$ and X-ray flux ratios of $F_{Aa}/F_{Ab}{\mathrm (core)} = 1.60$ and $F_{Aa}/F_{Ab}{\mathrm (wing)} = 1.29$.  The discrepant core and wing ratios suggest some level of contamination in the core extraction regions, assuming that the wing regions should have little, if any, contamination from source confusion.

Furthermore, there is some evidence that the core regions suffer from pileup. Specifically, using the CIAO tool \verb+pileup_map+) we find pileup rates of $\sim$15\% and $\sim$7\% for the core and wing extraction regions, respectively, for each component.  We caution that the \verb+pileup_map+ tool is not compatible with sub-pixel CCD event repositioning, and thus we are unable to establish with confidence which component suffered greater pileup.   

Given the likelihood of both photon pileup and cross-source contamination in the core extraction regions, we calculated the total X-ray flux (hence luminosity) of each component using the counts extracted from the wing extraction region, modified by a scaling factor that is a function of the area of the wing region and the effective encircled energy fraction of the annulus to which the flux is scaled. The shape of the wing extraction region was chosen to partially preserve the radial dependence of the e.e.f. when scaling the wing extraction flux.  The total X-ray flux ($F_X$) is calculated as $F_X \sim F_{Xwing} \times s.f.$ where the scale factor can be estimated as

\begin{equation}
\hspace{0.5in}s.f. =  \frac{\pi \hspace{0.5mm} (r_{0.8''})^{2} -\pi \hspace{0.5mm} (r_{0.25''})^{2}}{Area_{wing}} \times \frac{1}{e.e.f.(47\%)}
\end{equation}

Essentially, the scale factor converts the flux from the wing extraction region to estimate the flux located in an annulus centered on the source with inner and outer radii of 0.25$''$ and 0.8$''$, corresponding to radii of the core and wing extraction regions, respectively.   The encircled energy fraction corresponding to this annulus is the difference between the e.e.f. at 0.8$''$ (87\%) and the e.e.f. at 0.25$''$ (40\%).  We thereby calculate a factor of 4.1$\times$$F_{Xwing}$ to estimate the actual non-contaminated flux of each source using a wing area of 0.96 $arcsec^{2}$ and an effective e.e.f. of 47\%. This scaling factor can be applied to both components, since the extraction regions were of the same area and shape (Figure \ref{515A_chan}) and because the X-ray spectrum of each component was very similar (see Section 3.1.2). The resulting X-ray luminosities for components HBC 515 Aa and Ab are 6.48$\times$10$^{31}$ and 5.35$\times$ 10$^{31}$ erg s$^{-1}$, respectively.  The slightly enhanced luminosity of HBC 515Aa compared to HBC 515Ab is consistent with the alignment of the readout streak with HBC 515Aa, suggestive that this star is responsible for the generation of the artifact.  Assuming each HBC 515A component has $L_{bol}$ = 11.5 $L_{\odot}$ \citep{Reipurth2010},  we find fractional X-ray luminosities of $L_X$/$L_{bol}$ $\sim$ 1$\times$ 10$^{-3}$, typical of highly magnetically active, pre-MS stars \citep[][and ref. therein]{Kastner2016}.  

\subsubsection{Spectral analysis of individual components}

\begin{figure*}
\centering
\includegraphics[scale=0.5]{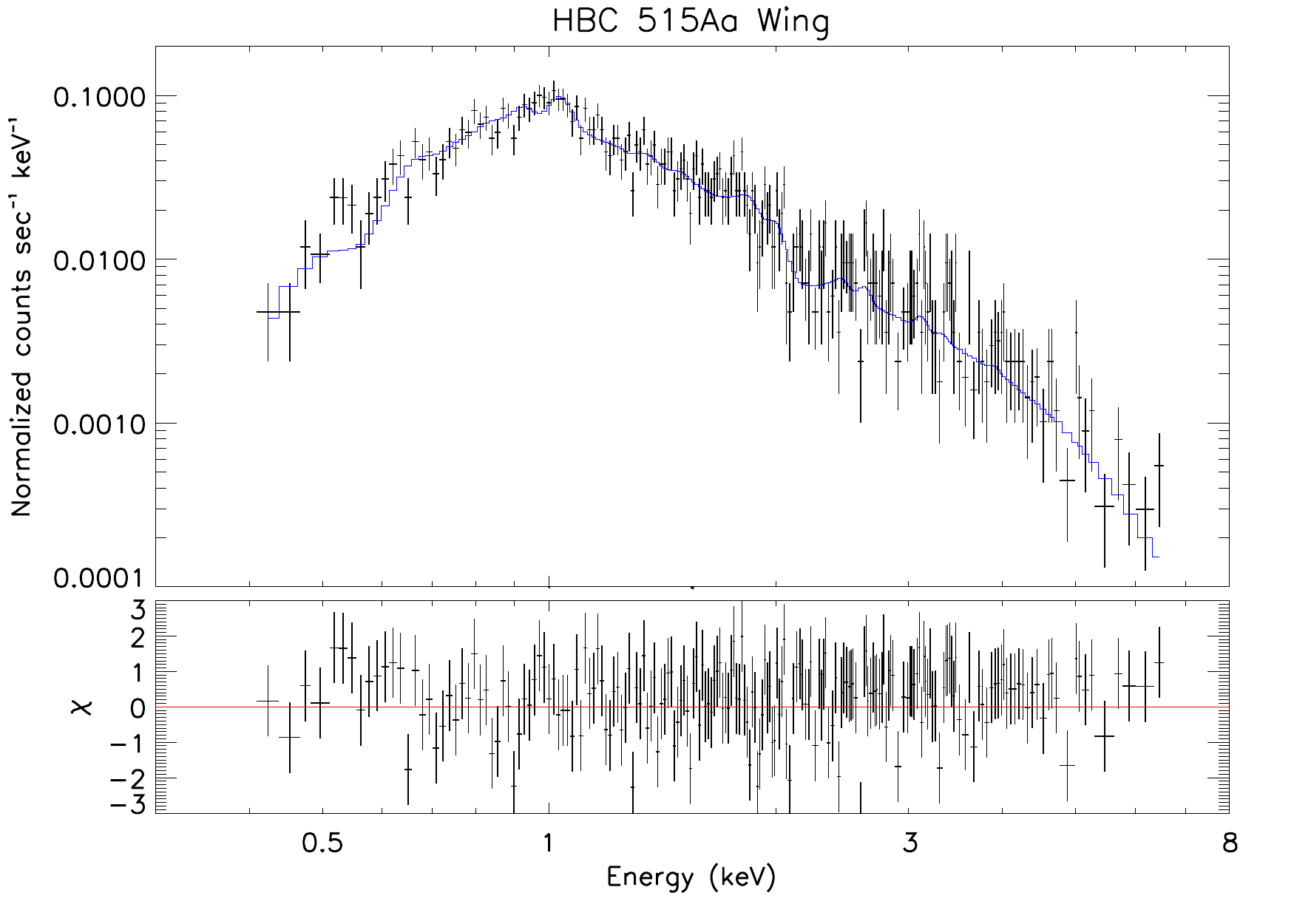}
\includegraphics[scale=0.49]{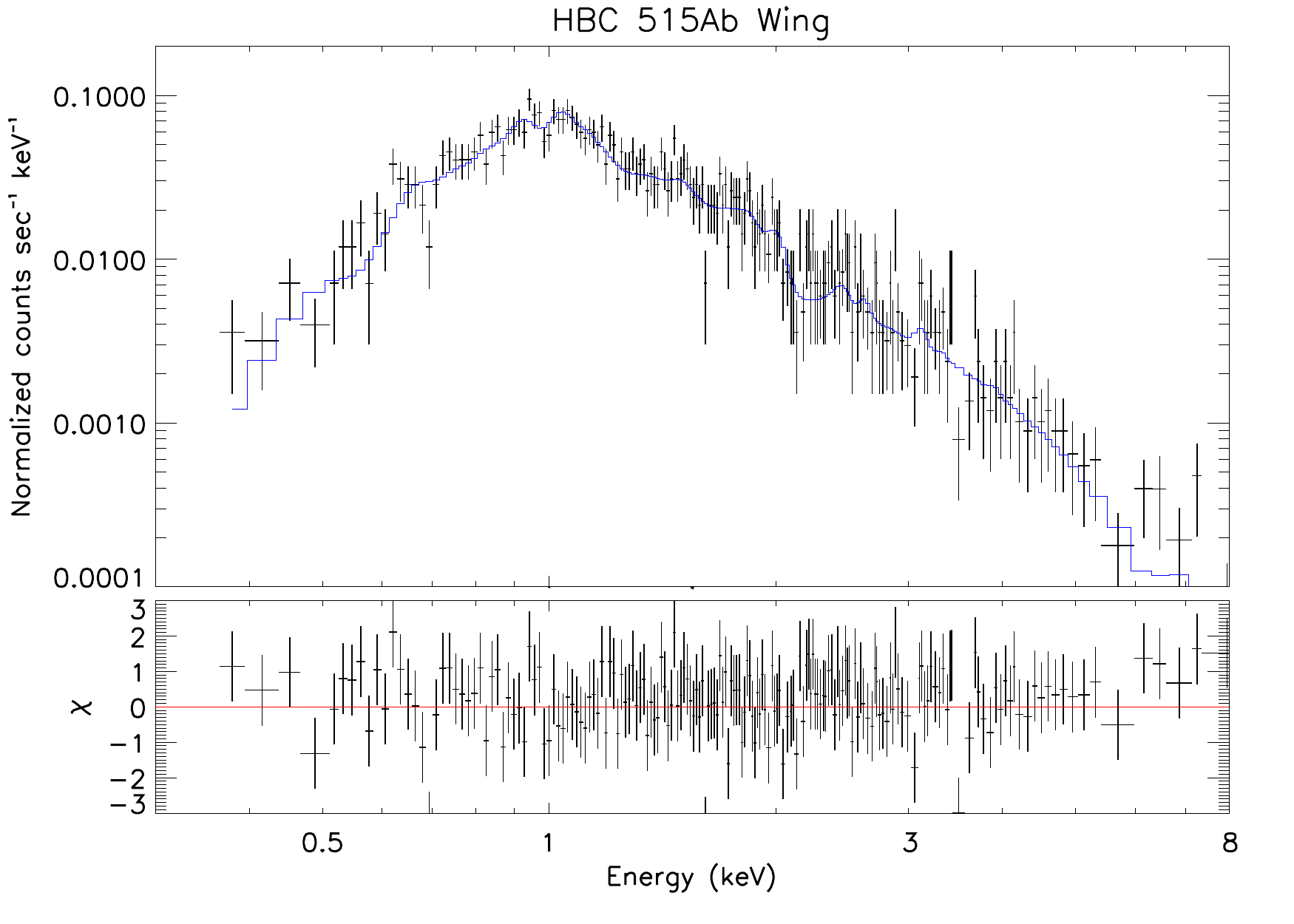}
\includegraphics[scale=0.5]{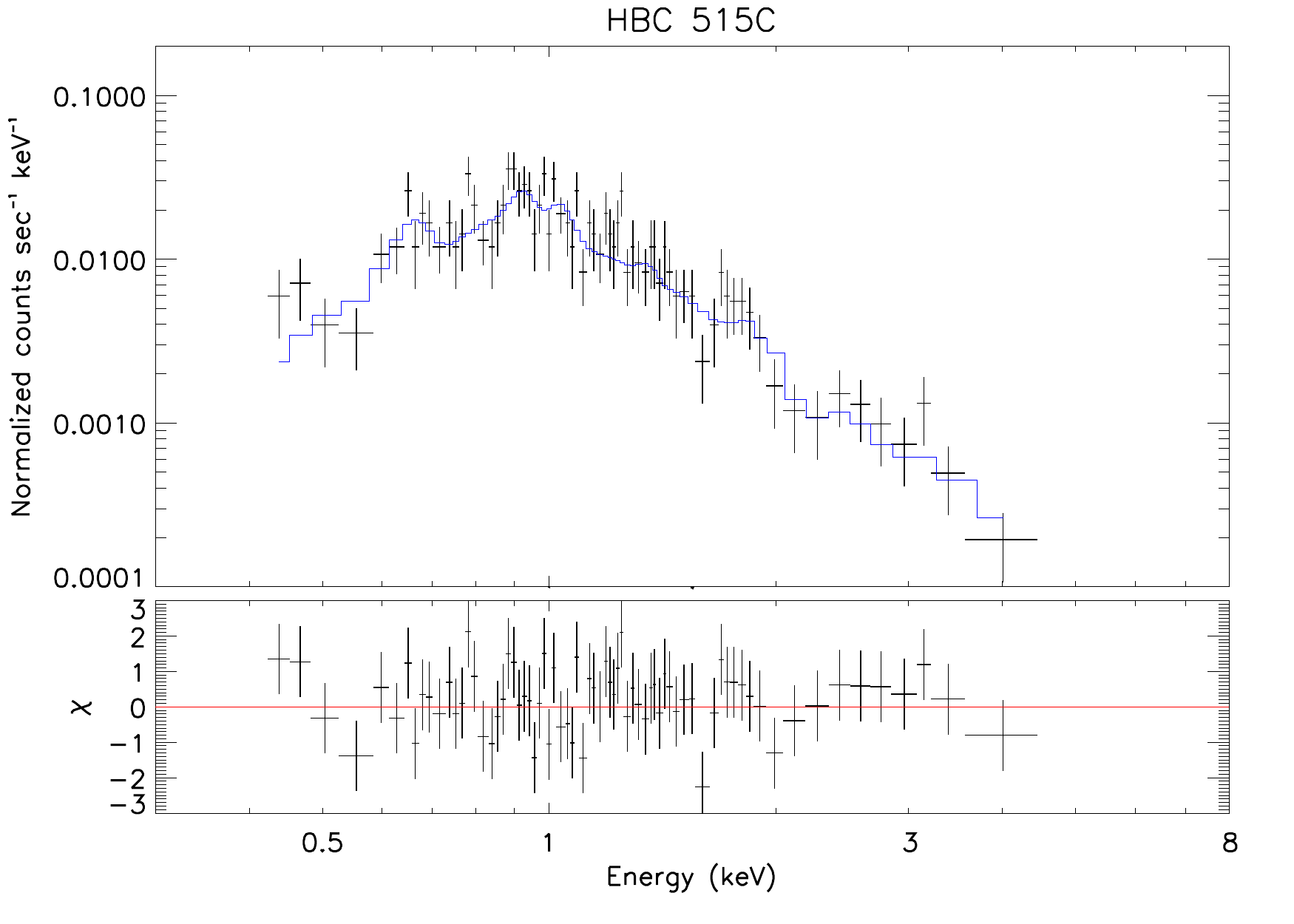}
\caption{X-ray spectra (black crosses), best fit models (blue) and residuals for the spectral extractions of HBC515 Aa, Ab and C.}
\label{spec}
\end{figure*}

The X-ray spectra extracted from the core and wing regions of HBC 515Aa and Ab were fit with an absorbed two-component thermal plasma emission model (vapec) assuming plasma abundances typical of T Tauri stars in Taurus \citep{Gudel2007,Scelsi2007}.  Modeling of low-resolution ACIS-S spectra is somewhat sensitive to the assumed Fe and Ne abundances \citep[e.g.,][]{Kastner2016}, so these abundances were left as free parameters in the fits. The same model was also used to fit the spectrum extracted for HBC 515C, which is the only other HBC 515 member with sufficient counts ($\sim$75) to warrant spectral fitting. The results of X-ray spectral fitting are listed in Table \ref{xspec_table} and displayed in Figure \ref{spec}. The X-ray spectra of components HBC 515 Aa and Ab, both of which peak near 1 keV, appear typical of X-ray-luminous pre-MS stars in Orion \citep[e.g.,][]{Principe2014}.  While spectra from both core and wing extraction regions can be fit reliably well (as evidenced by reduced $\chi^{2}\sim1$ in each case), we only list in Table \ref{xspec_table} the model parameters resulting from the fit to the spectrum yielded by the wing extraction region --- which suffers $\sim$50\% less pileup and less flux contamination from its nearby binary counterpart --- along with the corrected (scaled) source fluxes.  The fit results indicate that the X-ray-emitting plasmas of HBC 515Aa and Ab have very similar X-ray plasma properties (i.e., similar temperatures and similar Ne and Fe abundances).  

The best fit to the X-ray spectrum of the fainter HBC 515C source indicates plasma components of similar temperatures and abundances to those associated with HBC 515Aa and Ab. For all three components (HBC 515Aa, Ab, and C), we infer a moderate intervening X-ray (gas) absorbing column density, $N_H$ $\sim 2.5$ $\times$ $10^{21}$ cm$^{-2}$, which corresponds to optical extinction (due to intervening dust) of $A_V \sim 1.4$ based on standard gas-to-dust extinction conversion relations for the ISM \citep[i.e., $N_H = 1.8  \times 10^{21} A_V$ cm$^{-2}$ mag$^{-1}$;][]{Ryter1996}.  Given the large uncertainties in $N_H$ (Table \ref{xspec_table}), the foregoing value of $A_V$ is not inconsistent with those estimated in \citet{Kun2008} via $R_C$ and $I_C$ photometry, i.e., $A_V$ = 0.93, 2.45 and 3.24 for HBC 515A (Aa+Ab), C, and D, respectively.

A more rudimentary X-ray spectral analysis was performed for the moderately low count spectrum ($\sim$50 counts) of HBC 515B, which lies on the readout streak of Component A (Figure \ref{FOV}). 
To estimate an X-ray flux, we first estimated an absorbing column density ($N_H$) using previously established relationships of median energy to $N_H$ \citep{Feigelson2005, Principe2014}.  We find a median energy for HBC 515B of 4.3 keV in the 0.3-8.0 keV band, corresponding to $N_H \sim$ 1.3 $\times$ 10$^{23}$ cm$^{-2}$. This high level of extinction is consistent with B's large, positive hardness ratio, $HR$ = $(H-S)/(H+S) = 0.6$, where $H$ is the count rate in the 2.0-8.0 keV band and $S$ is the count rate in the  0.3-2.0 keV band. Furthermore, comparing the spectrum of HBC 515B with that of the readout streak, it is apparent that the 0.3-2.0 keV region of B's spectrum is heavily contaminated, such that the foregoing value of $N_H$ may be an underestimate. 

Since the readout streak contaminates the soft part of the spectrum, and most of the counts from HBC 515B originate in the hard ($>$ 2 keV) part of the spectrum, we estimated Component B's X-ray flux from its hard-band (2.0-8.0 keV) count rate (Table \ref{counts_table}). We used WebPIMMS (Portable, Interactive Multi-Mission Simulator) and the Cycle 12 response of Chandra ACIS-S, and adopted an absorbed ($N_H=1.3 \times 10^{23}$ cm$^{-2}$) apec plasma model with spectral parameters similar to those determined from spectral fitting of HBC 515C (i.e., $kT$ = 1.9 keV, abundance = 0.4 solar) to convert count rate to absorbed flux and intrinsic luminosity.  We find that B's count rate of 1.63 ks$^{-1}$ (Table \ref{counts_table}) corresponds to a 0.3-8.0 keV flux of $F_X=3.5\times10^{-14}$ erg s$^{-1}$ cm$^{-2}$ and an intrinsic X-ray luminosity of $L_X=8.0 \times 10^{30}$ erg s$^{-1}$, assuming a distance of 400 pc.  

Although considerably fainter than its HBC 515 companions, we also report a clear X-ray detection of HBC 515D.  Its median energy of 1.1 keV and hardness ratio of HR $= -1$ indicate very little X-ray absorption, similar to the spectral fitting results for HBC 515Aa, Ab, and C.  Adopting the same WebPIMMS plasma model with the same value of $N_H$ for HBC 515D as determined for HBC 515Aa, Ab, and C, we find that Component D's count rate of 0.574 ks$^{-1}$ corresponds to a flux of $F_X=5.2\times10^{-15}$ erg s$^{-1}$ cm$^{-2}$ and an intrinsic X-ray luminosity $L_X=9.9 \times 10^{28}$ erg s$^{-1}$.  

\subsubsection{X-ray light curves}

\begin{figure*}
\centering
\includegraphics[scale=0.46]{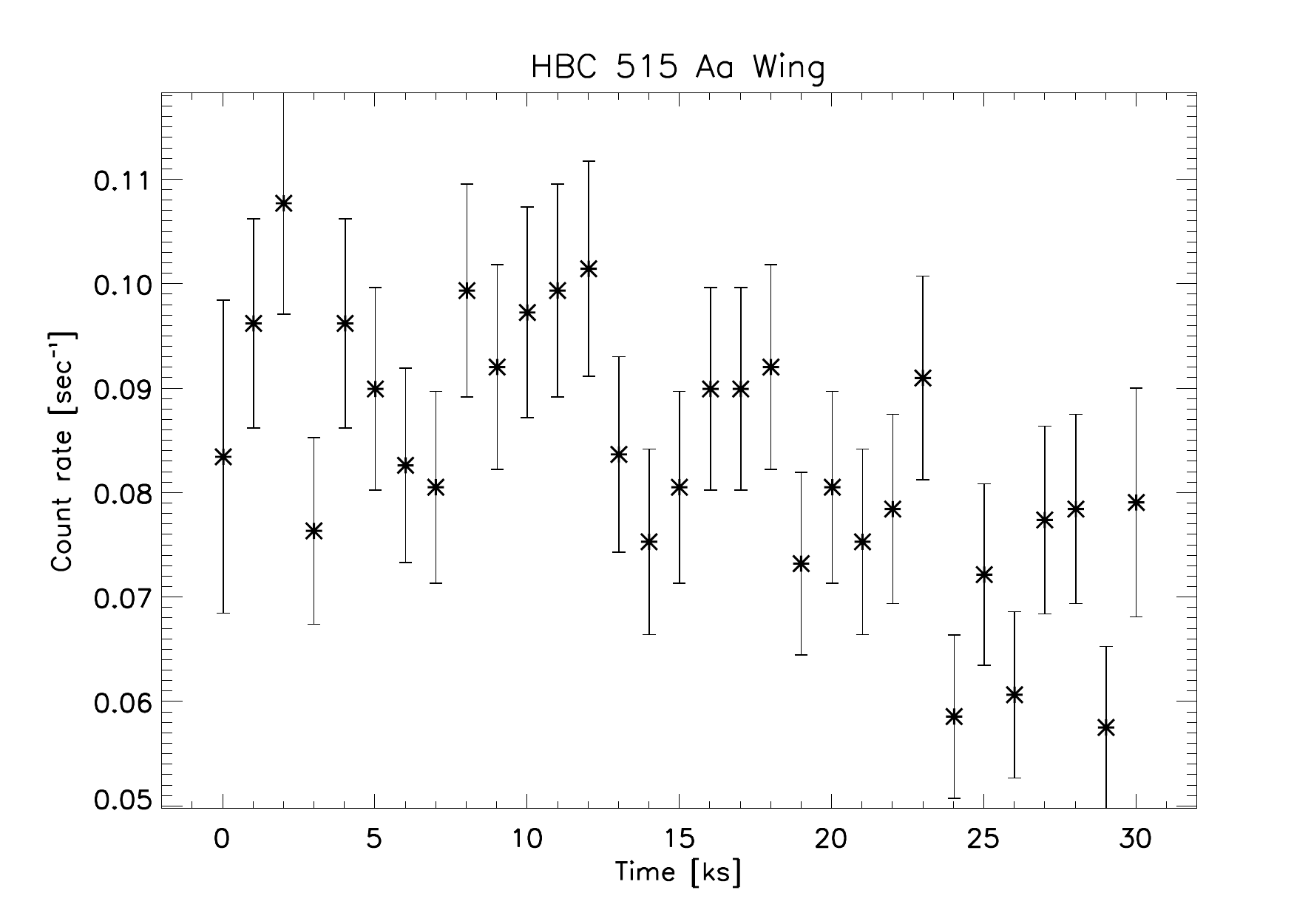}
\includegraphics[scale=0.46]{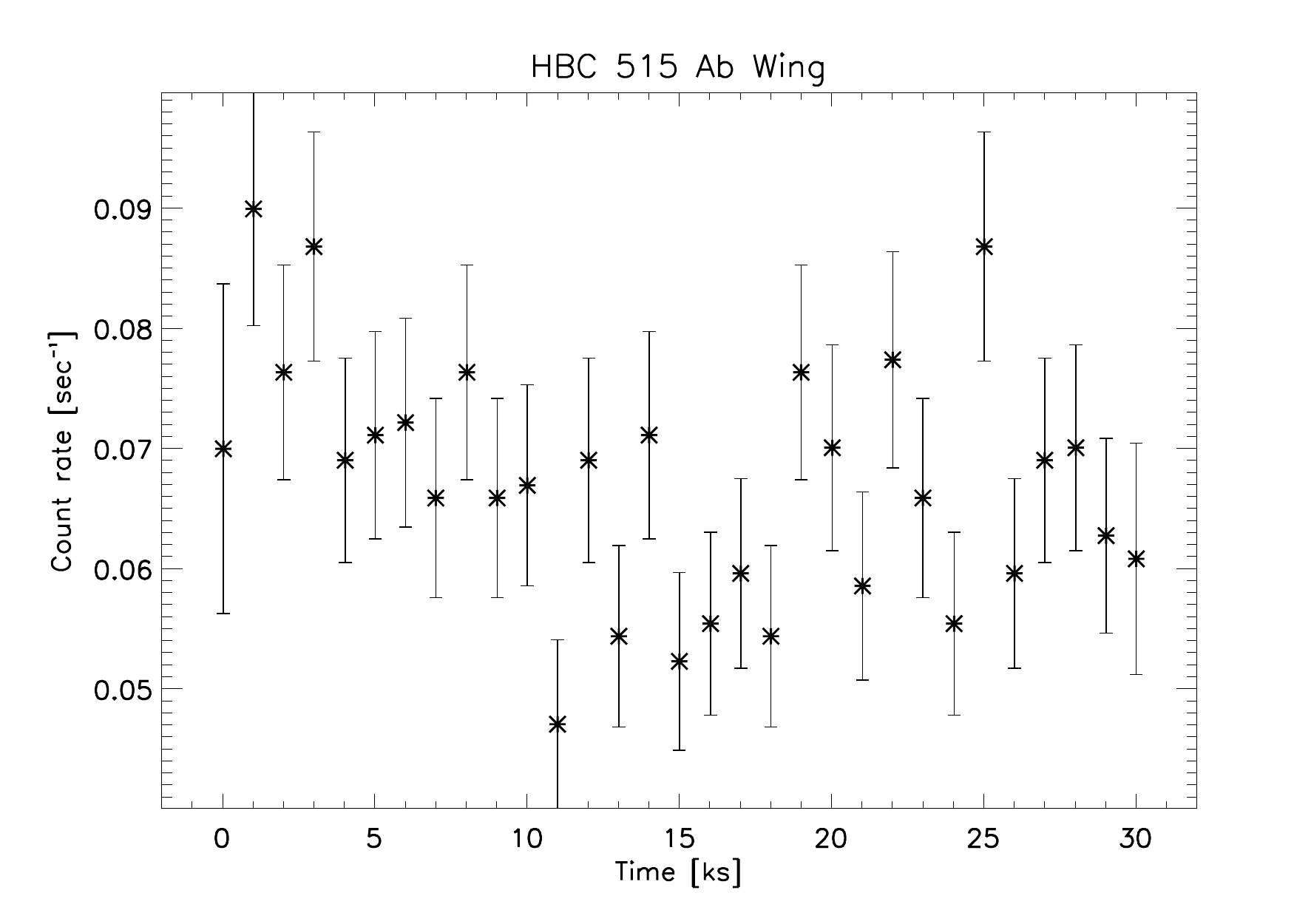}
\includegraphics[scale=0.46]{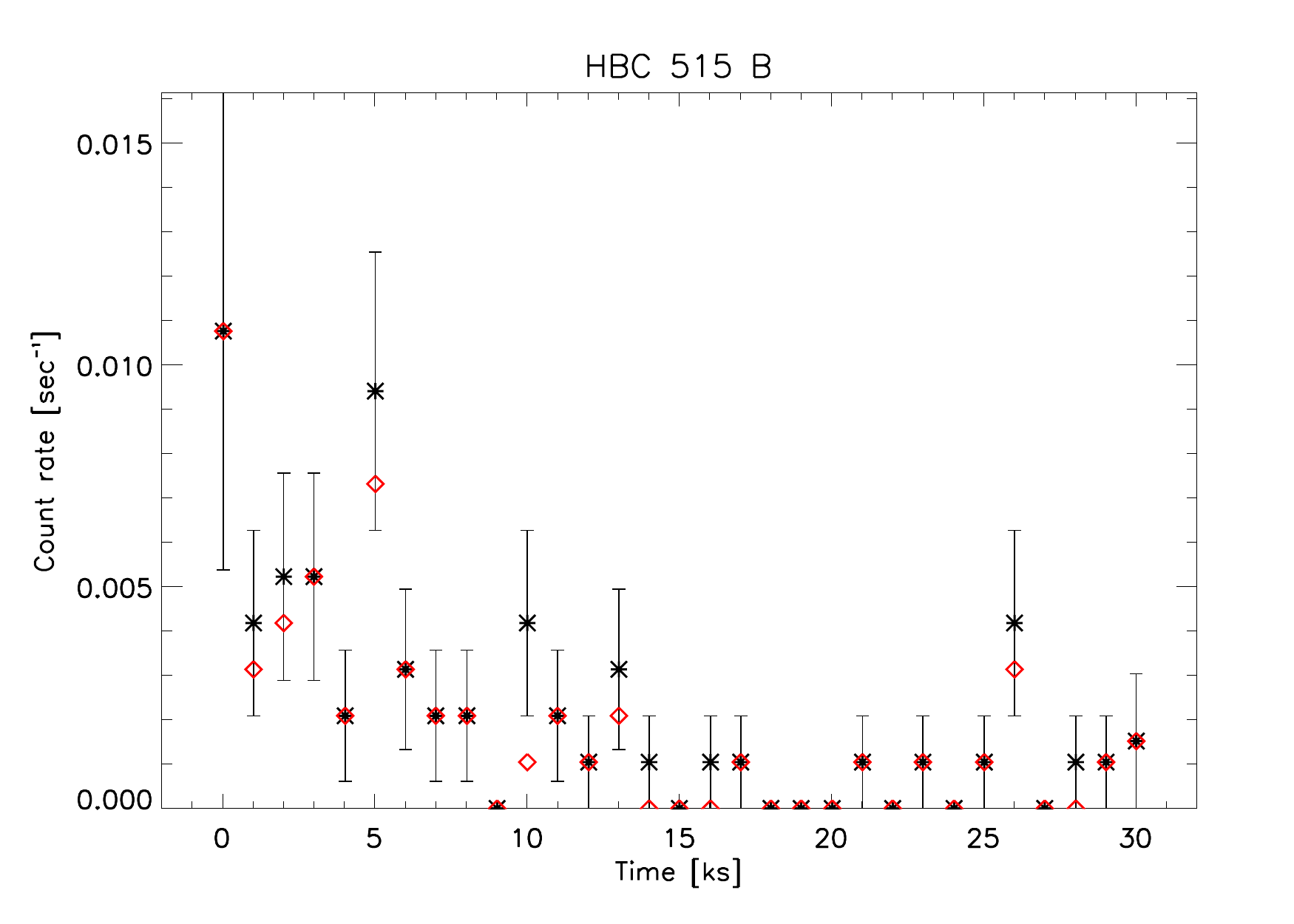}
\includegraphics[scale=0.46]{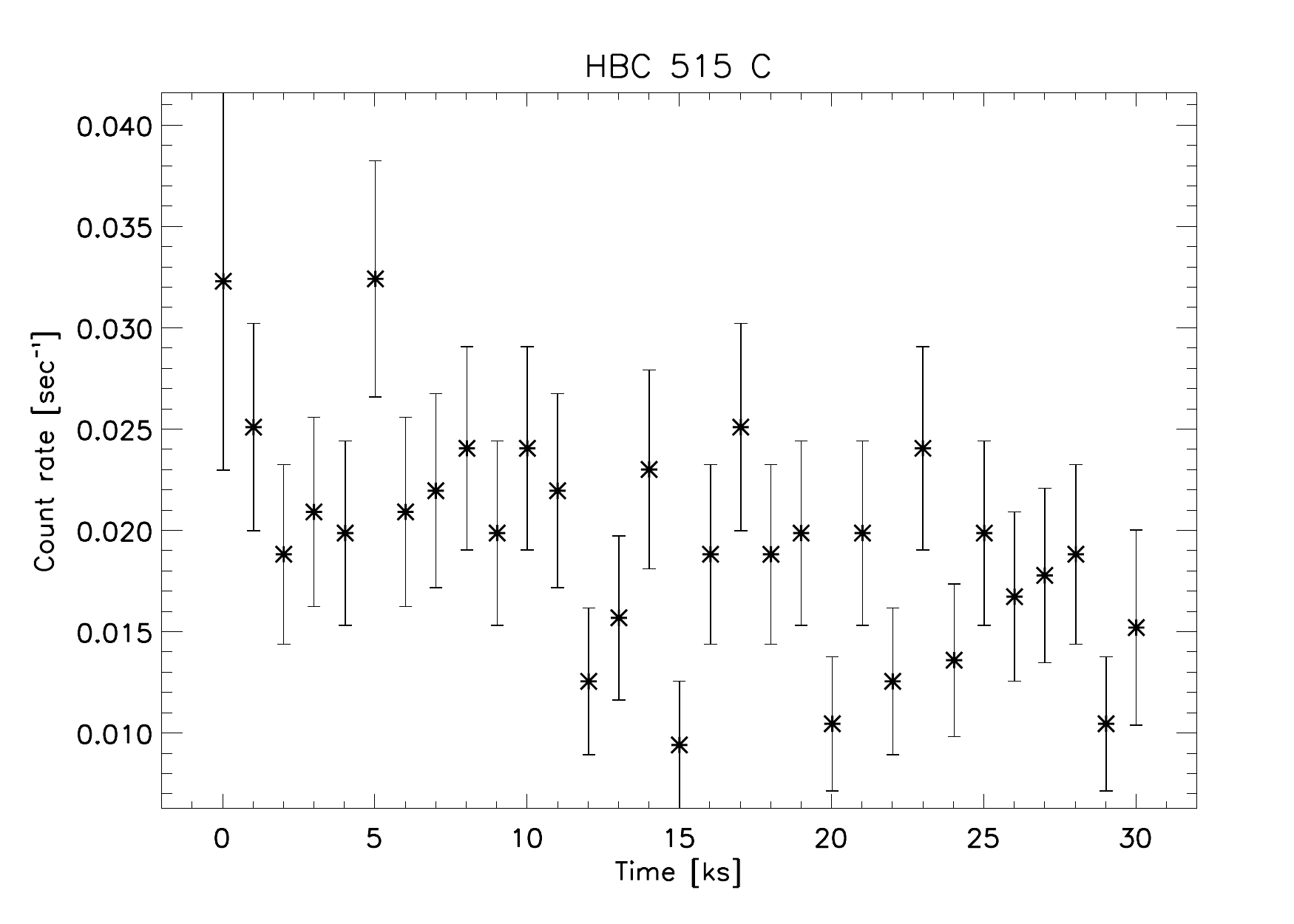}
\includegraphics[scale=0.46]{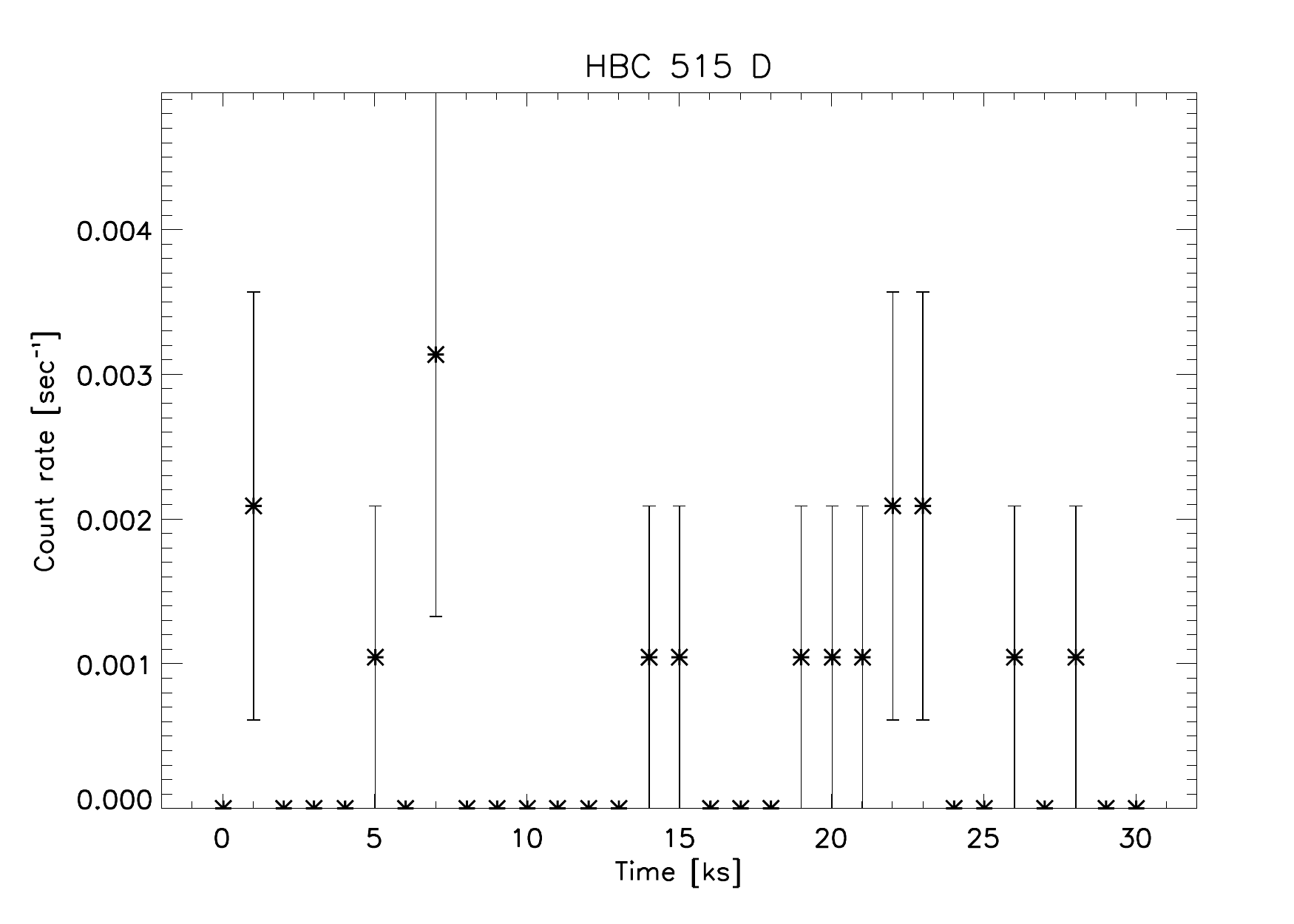}
\caption{ HBC 515 X-ray light curves during the course of the $\sim$29 ks observation with time bins of 1 ks and 1$\sigma$ count rate errors overlaid.  Time $t = 0$ corresponds to the start of the observation. Light curves for HBC 515B were extracted from the 0.3-8.0 keV band (black) and 2.0-8.0 keV band (red diamond) to highlight the 0.3-2.0 keV contamination from the readout streak (section 3.1.2)}.
\label{LC_515}
\end{figure*}

The light curve of each HBC 515 component is presented in Figure \ref{LC_515}. A variability test using CIAO \verb+glvary+, which utilizes the the Gregory-Loredo algorithm \citep{Gregory1992}, indicates little to no variability in the X-ray count rates of HBC 515Aa, Ab, C, and D.  In light of the hard median energy of HBC 515B (4.3 keV) the 2.0-8.0 keV light curve is shown in red in Figure \ref{LC_515} to indicate the level of contamination from the readout streak observed in the 0.3-8.0 keV bandpass. A flaring event was detected in the light curve of HBC 515B, indicating an approximate order of magnitude change in X-ray count rate $\sim$5ks into the observation. The \verb+glvary+ test indicated a variability probability greater than 99\% for both the broad and hard band light curves of HBC 515B.  

\subsection{SMA 1.3 mm continuum and CO(2--1) imaging}

\begin{figure*}
\centering
\includegraphics[scale=0.35]{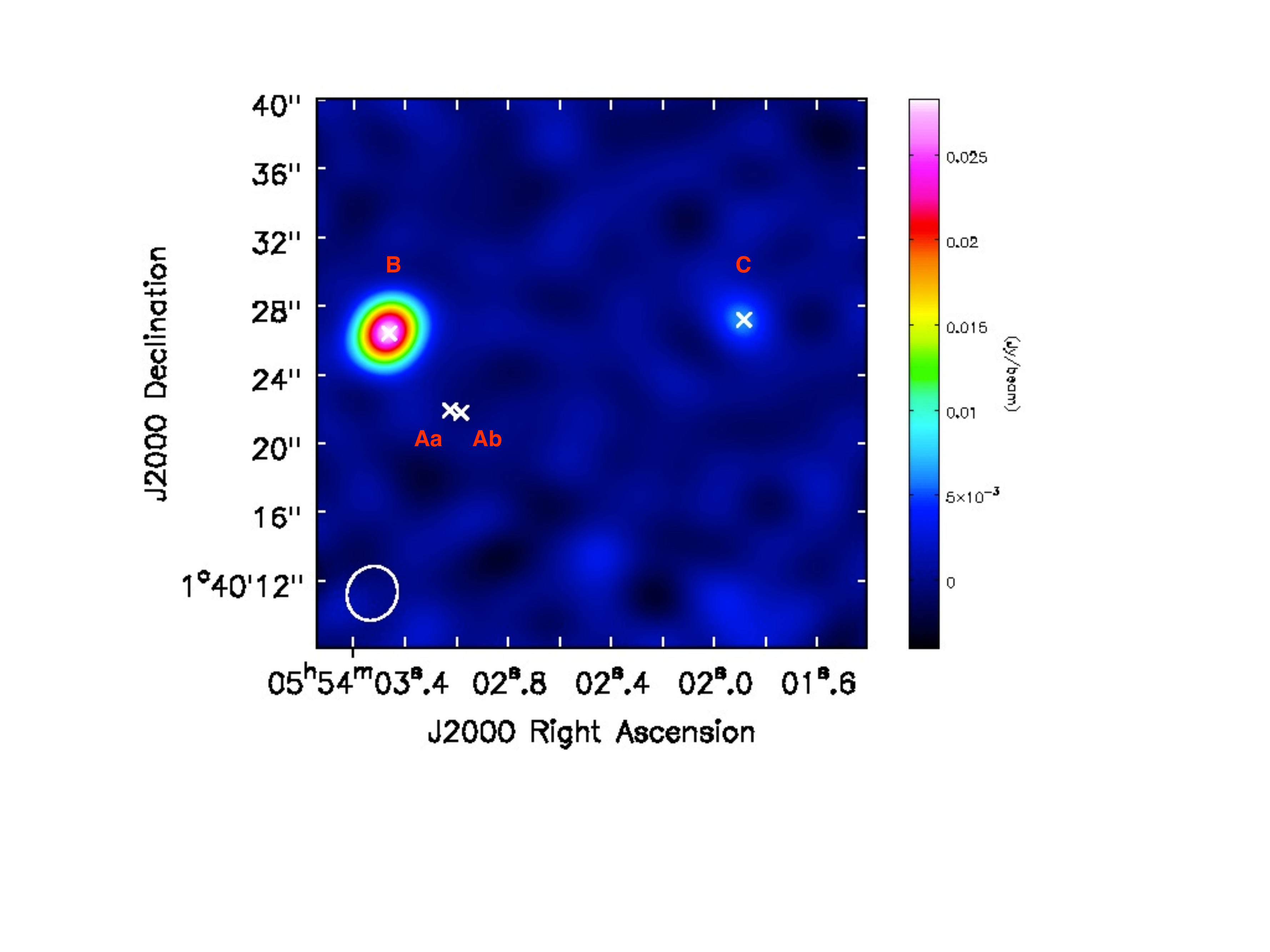}
\includegraphics[scale=0.345]{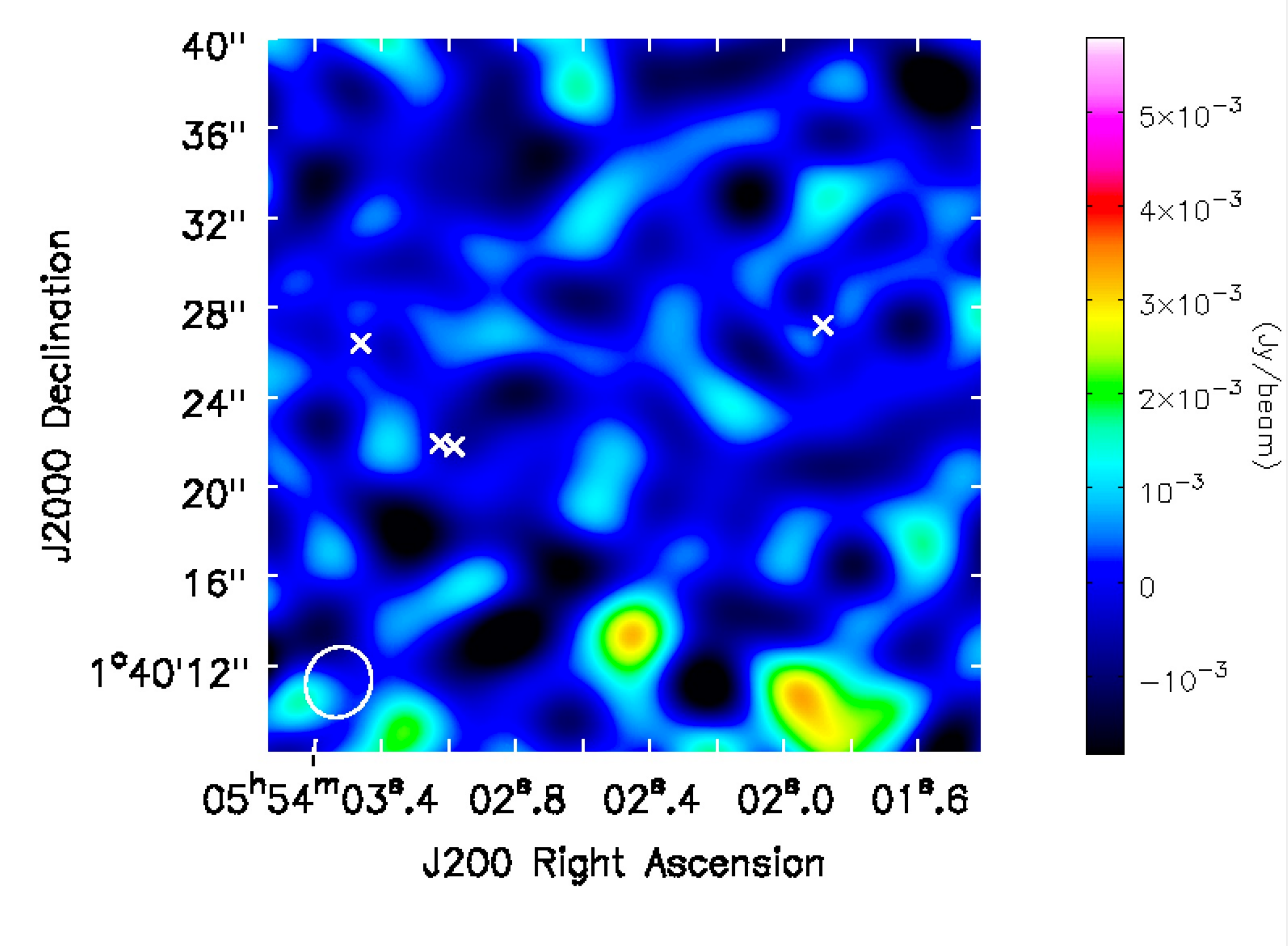}
\caption{SMA 1.3mm dust continuum image of the HBC 515 system (Aa+Ab, B and C; left) and residuals from 2D gaussian fits of HBC 515B and C (right) where the position of each member is identified with white crosses.  The beam size is shown in the lower left. }
\label{sma}
\end{figure*}

The image obtained from the SMA 230.538 GHz (1.3 mm) continuum observation of HBC 515 is presented in Figure \ref{sma}. This image reveals that a pair of mm-wave continuum emission sources are detected, coincident with the locations of HBC 515 B and C.  No continuum emission is detected in the vicinities of HBC 515A or HBC 515D, although we caution that the latter lies beyond the half-power radius of the SMA observation and thus requires a correction factor to determine its upper limit.  Given a radial offset of $\sim$30$''$, the noise at the location of HBC 515D is increased by a factor of 2.7 and thus its 3$\sigma$ detection upper-limit has been scaled accordingly.  Fluxes (and flux upper-limits) determined for the 1.3 mm emission sources associated with HBC 515 members are listed in Table \ref{sma_table}.

Millimeter emission from circumstellar material can be converted to a circumstellar disk dust mass by assuming the emission is thermal and optically thin, where the latter condition may not be met in the innermost regions of evolved disks (Williams \& Cieza 2011).  The dust mass can then be calculated via
\begin{equation}
\hspace{1.2in}M_{dust}= \frac{F_\nu d^2}{\kappa_\nu B_\nu(T)},
\end{equation}
where $B_\nu(T)=2\nu^2kT/c^2$ represents Rayleigh-Jeans (blackbody)  emission from dust at a temperature $T$ and $\kappa_\nu$ is the dust opacity, which is a power-law function of submm frequency, i.e., $\kappa_\nu=0.1(\nu/10^{12})^\beta$ \citep{Beckwith1990}.  The power law index $\beta$ is related to the size distribution and composition of the dust particles and is typically $\sim$1 for circumstellar disks \citep{Williams2011}.  For a dust temperature $T=20$ K, distance d = 400 pc and $\kappa_\nu$ = 1.8 cm$^{2}$ g$^{-1}$, we thereby estimate disk dust masses for HBC 515B and 515C of 139.5 M$_E$ and 30.4 M$_E$, respectively.  If we assume a gas to dust mass ratio of 100, the implied total disk masses are 44.3  M$_J$ and 9.64 M$_J$, respectively. Based on the sensitivity of the continuum observation, we estimate a 3$\sigma$ upper limit of 2.86 mJy, corresponding to a dust upper-limit disk mass of 13.4 M$_E$, for any non-detected disks around HBC 515 Aa and HBC 515Ab.  Due to the radial offset of HBC 515D, we estimate a 3$\sigma$ upper limit of 7.7 mJy, corresponding to a dust upper-limit disk mass of 36 M$_E$

Moment 0 integrated intensity maps extracted from the SMA $^{12}$CO and $^{13}$CO $J = 2\rightarrow1$ spectral line mapping data for the HBC 515 field reveal several sources; however, none of these are clearly associated with any of the HBC 515 members, with the possible exception of HBC 515B. Specifically, a $^{12}$CO emission source with integrated flux of $7.09 \pm 0.4$ Jy km s$^{-1}$ is detected within 1$''$ of the centroid of HBC 515B's continuum emission. However, its narrow emission line profile suggests this source is most likely molecular cloud material, as opposed to circumstellar emission. Similarly, moment 1 velocity maps generated from the SMA $^{12}$CO and $^{13}$CO data do not yield double-peaked (Keplerian) velocity profiles for any of the other CO sources in the field.   We estimate a $3\sigma$ upper limit of 1.2 Jy km s$^-1$ for the $^{12}$CO intensity for the nondetected members HBC 515Aa, HBC 515Ab, HBC515 B and HBC 515C, based on the sensitivity of the $^{12}$CO data.  All upper limits are listed in Table \ref{sma_table}.

  \section{Discussion}

\subsection{ HBC 515A: an X-ray luminous, intermediate-mass, pre-MS binary}

The Chandra imaging (\S 3.1) demonstrates that HBC 515A dominates the total X-ray flux from the HBC 515 system, and furthermore establishes this YSO as one of the most X-ray-luminous late-type stars in the Orion star-forming region \citep[see, e.g.,][]{Preibisch2005,Principe2014}. These results hence place HBC 515A among the rare examples of relatively massive, X-ray luminous ``weak-lined T Tauri stars'' that are binaries --- in many cases, hierarchical binaries --- and that have shed their disks at very early stages of pre-MS evolution. Such systems potentially include \object{DoAr 21} (see below), \object{HD 283447}  \citep[= V773 Tau;][]{Tsuboi1998,Torres2012}, HDE 283572 \citep{Kenyon1995,Gudel2007}, V410 Tau \citep{Stelzer2003}, V826 Tau \citep{Giardino2006}, HD 155555  \citep{Strassmeier2000,Lalitha2015}, and perhaps HBC 502 \citep{Simon2004,Principe2014}.

Among the objects just listed, DoAr 21 may be the most similar to HBC 515A in terms of its main system characteristics, i.e., binarity, component masses, spectral types, and ages. 
DoAr 21 is an unusually X-ray bright ($\sim 10^{32}$ erg s$^{-1}$) wTTS binary \citep[separation 1-2 AU;][]{Loinard2008} in the Ophiuchus Molecular Cloud with a spectral type between K0-K2, age of $\sim$0.8 Myrs, and component masses of $\sim$1.8 M$_\odot$. Like HBC 515A, it furthermore also exhibits a modest 24 micron excess, unusual for wTTSs,  while showing no indication of near-IR excess at shorter wavelengths.  \citet{Jensen2009} found that the mid-infrared excess is most likely associated with a small-scale photodissociation region (PDR) located $\sim$100's of AU from the star that is perhaps excited by the UV emission of DoAr 21.  Such a scenario may also explain the 24 $\mu$m excess seen in HBC 515A, given the lack of evidence of an orbiting disk in the near-IR \citep{Reipurth2010} and at 1.3 mm (this work).

As in the case of HBC 515A, DoAr 21 displays an X-ray spectrum typical of pre-MS stars in Orion \citep{Preibisch2005} that is characterized by plasma components with temperatures of $\sim$1 and 3 keV.  Unlike HBC 515A, which displayed little to no variability during our $\sim$30 ks Chandra observation (Fig. \ref{LC_515}), flares were detected during two independent $\sim$100 ks observations of DoAr 21.  One epoch displays small-scale flares superimposed on a slowing declining light curve \citep{Gagne2004} whereas the more recent observations displayed an impulsive flare during which the temperature of the X-ray emitting plasma increased by a factor of $\sim$2 \citep{Jensen2009}.  Indeed,  \citet{Jensen2009} concluded that DoAr 21 exhibits X-ray flares at a rate of nearly one per day. This high rate may reflect the increased likelihood of flaring in small-separation TTS binary systems, wherein the coronal activity of each component may be influenced by the interactions of the stars' magnetospheres \citep{Stelzer2000}.  While the physical separation of HBC 515Aa and 515Ab is too large ($\sim$200 AU) to have interacting magnetospheres, HBC 515 Ab may itself be a close binary \citep{Reipurth2010}, and a longer X-ray observation may detect flaring due to such interactions.

V773 Tau is another example of an extremely X-ray-bright ($L_X$ $\sim$ 10$^{31}$ erg s$^{-1}$), massive, multiple-component system.  Like DoAr 21, V773 Tau has been observed to undergo  flaring events during which the plasma reaches temperatures higher than kT $\sim$ 10 keV \citep{Tsuboi1998}.  Like HBC 515, the multiple components of the $\sim$5 Myr-old V773 Tau system exhibit a range of evolutionary states: V773 Tau ABC is a $\sim$ 3.7 $\pm$ 0.7 $M_{\odot}$ triple system consisting of two wTTS spectroscopic binary (SB) components (AB, K2+K5) with a separation of 0.3 AU, and a companion, V773 Tau C, that is a cTTS and is separated from AB by $\sim$15 AU \citep[$\sim$0.1$''$ assuming a distance of 148 pc;][]{Ghez1993,Leinert1993,Welty1995,Duchene2003,Woitas2003}. This triple system was initially thought to include an evolved, "fossil" protoplanetary disk on the basis of its mid-IR excess. However, more recent high angular resolution near-IR and optical observations attribute some of the system's IR excess to both V773 Tau C and a more recently identified fourth component, V773 Tau D, which is an infrared companion located at a projected distance of $\sim$30 AU from V773 Tau C \citep{Duchene2003}.

The largest angular separation between any component in this system is $\sim$0.5$''$, very similar to the angular separation of HBC 515Aa and Ab.  Moreover, if analogous to V773 Tau, the unusual 24 micron excess detected in HBC 515A (Aa and Ab) may be attributed to a heretofore undetected companion to Aa or Ab. This would be consistent with the suggestion by \citet{Reipurth2010} that the PSF of component Ab was slightly elongated, potentially indicating the existence of an unresolved, additional component. However, if there is an unresolved additional component with a circumstellar disk then it must be less massive than $\sim$ 4 M$_{Jup}$, given its non-detection in the 1.3 mm SMA observations presented here.   

Objects like HBC 515A, DoAr 21, V773 Tau and others mentioned in the beginning of this section are of interest to studies of disk longevity, given that stellar multiplicity and X-rays likely both play an important role in the formation and survival of circumstellar disks. With regard to binarity, \citet{Kraus2012} find that 2/3 of all close ($\leq$ 40AU) binaries in the 1-2 Myr Taurus-Auriga star-forming region do not host circumstellar disks, whereas $\sim$80-90 \% of wide binaries and single stars retain their disks for at least 1-2 Myr.  This suggests that if binary separation alone were responsible for disk formation and/or longevity, HBC 515Aa and Ab would likely still harbor circumstellar disks, given their projected separation of 200 AU. If each of the HBC 515A components are in fact multiple systems with small separations, then this could account for the absence of circumstellar material in these systems.

Given the high X-ray luminosities of the HBC 515A, DoAr 21 and V773 Tau systems, however, it is likely X-ray photoevaporation plays a significant role in the dispersal of their disks \citep{Gorti2009,Owen2012}.  The mass loss rate of each HBC 515A component that is due only to photoevaporation of disk material from stellar X-ray-induced disk gas photoevaporation (i.e., not accounting for viscous accretion onto the star), $\dot{M_{X}}$, can be estimated as $\dot{M_{X}}$ = 8 $\times$ 10$^{-9}$ $L_{X30}$ M$_{\odot}$ yr$^{-1}$ \citep{Owen2012}. We caution that this $\dot{M}_X$ vs $L_X$ relation was obtained from a model that was developed for the solar-mass regime and that the photoevaporative mass loss rate is sensitive to the irradiating stellar spectrum \citep[e.g.,][]{Gorti2015}. Nonetheless, adopting the \citet{Owen2012} $\dot{M}_X$ vs $L_X$ relation and making the additional assumptions that the the X-ray luminosity of each component as well as the star-disk absorbing columns have remained constant during the course of the $\sim$0.5 Myr system lifetime, we obtain an estimate of $\sim$ 250 M$_{Jup}$ for the mass lost from the disk of each component due to X-ray-induced photoevaporation. This effect may be enhanced when considering the X-ray luminosity of each component on the other component's disk. We note that although strong X-ray flares may have been present throughout the prior evolution of HBC 515, the flare timescale is likely too short to result in a change in photoevaporative flow \citep{Owen2011b}.  

\subsection{HBC 515B: a millimeter-luminous protostar}

The detection of a 1.3mm emission source coincident with HBC 515B is consistent with the IR-SED-based evidence suggesting that this object is an embedded protostar \citep{Reipurth2010}. Moreover, our 1.3mm based estimate of a mass of 44 M$_{jup}$ for the disk and/or envelope of HBC 515B is similar to the disk masses of protostars in Taurus-Auriga. Specifically, \citet{Andrews2005} found a median disk mass of 31.4 M$_{Jup}$ for 16 Class I protostars in Tau-Aur, compared to a median disk mass of 3.14 M$_{Jup}$ for 64 Class II stars in the same region.  This comparison confirms that HBC 515B is the least evolved of the HBC 515 members, and suggests that it is most appropriately identified as Class I.  

The detection of $^{12}$CO gas emission at the location of HBC 515B whose narrow emission line profile is indicative of ambient cloud emission furthermore suggests this object is still located deep in the molecular cloud environment. This notion is further substantiated by the large $N_H$ inferred from its median X-ray energy. Moreover, the significant X-ray luminosity, absorption, and variability of HBC 515B are all consistent with those of Class I objects in L1630, another star-forming region in Orion whose variability and X-ray properties are well characterized \citep{Principe2014}.  The approximate order of magnitude change in X-ray count rate exhibited by HBC 515B during the course of these Chandra observations may hence be associated with a coronal flare or a large flare from a star-disk magnetic reconnection event. 

\subsection{HBC 515C and 515D }

HBC 515C exhibits an X-ray spectrum similar to that of HBC 515 Aa and Ab, even though it is significantly less massive and less X-ray luminous  (Table \ref{xspec_table}).  Their similar levels of X-ray absorption and the soft (1.1 keV) median energy of HBC 515D furthermore support the notion that HBC 515Aa, Ab, C, and D likely have formed in close proximity.  HBC 515C exhibits a weak but detectable level of 1.3 mm emission, and our estimate of a total disk mass of $\sim$ 10 M$_{Jup}$ within its circumstellar disk supports the conclusion, based on its IR SED, that this object is a pre-MS star with a transition disk. While similar in mass to HBC 515C, HBC 515D was not detected in the 1.3mm SMA continuum observations.  Given HBC 515D's status as a Class II YSO,  the non-detection of 1.3mm disk emission suggests that either the disk is void of large grains or its true brightness is very near the sensitivity of the observations.

\section{Summary and Conclusions}

The Chandra X-ray and SMA mm-wave imaging of the HBC 515 system presented here underscores the diverse stages of pre-MS stellar evolution represented by its five, presumably coeval member stars. We find that HBC 515Aa and Ab comprise a highly X-ray luminous wTTS binary, with $L_X$ = 6.5 and 5.4 $\times$ 10$^{31}$ erg s$^{-1}$, respectively. The HBC 515A binary displays no 1.3 mm emission associated with circumstellar disk material, despite its excess 24 $\mu$m emission.  We speculate that the spatially unresolved 24 $\mu$m emission may originate from surrounding cloud material that is either a remnant of the star formation process or potentially is associated with a small-scale PDR that is excited by (as-yet-undetected) UV emission from HBC 515A. Moreover, the absence of circumstellar disk and/or envelope of mass around HBC 515 Aa and Ab are likely the result of a combination of factors, including intense X-ray irradiation and potential, multiple-component binaries.   The similarity of HBC 515A to the other (fairly rare) examples of diskless, X-ray-luminous ``wTTS,'' many of which are binaries, further suggests that binarity is a ``wild card'' that can strongly influence --- in these cases, severely truncate --- disk lifetimes.  Adaptive optics imaging at resolutions of $\sim$ 0.1 arcseconds in the near- and mid-IR would be required to determine the the number of stellar components in the 515 Aa/Ab system and the origin of its excess 24 micron emission.

Our detection of a mm-wave continuum source associated with HBC 515B, along with our inference of a large X-ray absorbing column ($N_H$ = 1.3 $\times$ 10$^{23}$ cm$^{-2}$) and the apparent detection of a strong X-ray flare, support the conclusion \citep{Reipurth2010} that this object is a deeply embedded Class I protostar. The 1.3mm dust continuum detection indicates the presence of a circumstellar disk of mass $\sim$44 M$_{Jup}$, and the $^{12}$CO data indicate that molecular cloud material lies in close proximity, consistent with its being the most deeply embedded, least evolved member of HBC 515. HBC 515C is an M4 pre-MS star with X-ray spectral properties similar to that of HBC 515Aa and HBC 515Ab, and HBC 515C hosts what is likely a transition disk with a mass of $\sim$10 M$_{Jup}$.  Their similar levels of X-ray extinction support the hypothesis that the members of HBC 515 originated from the same parent molecular cloud.  HBC 515D is an M3 cTTS with an X-ray luminosity an order of magnitude fainter than HBC 515C, likely the result of the X-ray flare associated with HBC 515C.  We report no detection of 1.3mm continuum emission at the location of HBC 515D, which indicates its disk mass is less than 11.4 M$_{Jup}$ based on the sensitivity of the observation. 

The diverse stellar evolutionary stages of HBC 515 components are important when considering the evolution of the HBC 515 system as a whole. If these stars are a single, presumably coeval multiple YSO system then the disparity between evolutionary state reflects the influence of pre-MS star mass, binarity, and X-ray luminosity in regulating the lifetimes of circumstellar, planet-forming disks and the timescales of star-disk interactions. If all five components initially formed from the collapse of the same parent cloud core, then the location of HBC 515B, C, and D suggests these components were dynamically removed from the system during the early stages of formation and thus, were able to retain their circumstellar disks (at least in the case of HBC B and C) due to their location away from the intense stellar X-ray and UV fields of HBC 515A.

\begin{acknowledgements}
      This research was supported via award No. GO1-12027X to RIT issued by the Chandra X-ray Observatory Center, which is operated by the Smithsonian Astrophysical Observatory for and on behalf of NASA under contract NAS8-03060.  DP acknowledges a CONICYT-FONDECYT award (grant 3150550) and support from the Millennium Science Initiative (Chilean Ministry of Economy; grant Nucleus RC 130007).  The Submillimeter Array is a joint project between the Smithsonian Astrophysical Observatory and the Academia Sinica Institute of Astronomy and Astrophysics and is funded by the Smithsonian Institution and the Academia Sinica.
\end{acknowledgements}

\begin{table*}
\centering
\caption{X-ray Sources and Classification}

\input{table_counts_v3.tex}
\footnotemark[1]{Measured in the 2.0 - 8.0 keV band to avoid contamination from the readout streak.}
\footnotemark[2]{\citet{Kun2008}.}
\footnotemark[3]{\citet{Reipurth2010}}
\label{counts_table}
\end{table*}

\begin{table*}
\centering
\caption{X-ray Spectral Fit Parameters with 2$\sigma$ Errors  \vspace{2.mm}}
\input{table_xspec_fit_v6.tex}
\hspace{2.2in}
\footnotemark[1]{Abundance relative to Solar \citep{Anders1989}. }
\footnotemark[2]{X-ray flux and luminosity were measured in the 0.3-8.0 keV X-ray band.}
\footnotemark[3]{Intrinsic $L_{X}$ was estimated from the wing extraction flux and scaled to account for the entire PSF of the source (see section 3.1).}
\label{xspec_table}
\end{table*}

\begin{table*}
\centering
\caption{SMA Fit Parameters with upper limits determined from the sensitivity of the observation. }
\input{table_SMA_hbc515_v3.tex}
\label{sma_table}
\hspace{1.1in}\footnotemark[1]{Upper limits were estimated by scaling the noise by a factor of 2.7 due to the location of HBC 515D (see Section 3.2).}
\end{table*}

%%%%%%%%%TABLES END %%%%%%%%%%%%%%

%%%%%%%%%FIGURES START %%%%%%%%%%%%%%

%%%%%%%%%FIGURES END %%%%%%%%%%%%%%


%\listofobjects

\end{document}

%% file: table_counts_v3.tex
%NOTE, version two of this table has a new count rate for 515 Aa because I redid the x-ray analysis with an aperture size of 0.25'' to match Ab.
%v3 includes the addition of YSO class and spectral type for ref

{\renewcommand{\arraystretch}{1.5}

\resizebox{\textwidth}{!}{
\begin{tabular}{ c  c  c  c  c  c  c  c }
%\begin{tabular}{| c | c | c | c | c | c | c | c |}
\hline

 & \object{HBC 515Aa} & HBC 515Aa Wing & \object{HBC 515 Ab} & HBC 515 Ab Wing & HBC 515 B & HBC 515 C  & HBC 515 D \\ \hline
 Right Ascension & 5:54:03.02 & - & 5:54:02.98 & - & 5:54:03.26 & 5:54:01.88  & 5:54:02.00 \\
Declination & +1:40:21.99 & - & +1:40:21.87 & -  & +1:40:26.48 & +1:40:27.22  & +1:40:55.51 \\
C. Rate [ks$^{-1}$] & 121.7$\pm{2.48}$ & 83.7$\pm{1.71}$ & 80.7$\pm{1.67}$ & 67.3$\pm{1.53}$ & \footnote{1}1.63$\pm{0.238}$ & 19.2$\pm{0.817}$  & 0.574$\pm{0.143}$ \\ 
IR Class\footnote{2} \footnote{3} & Class III & -- & Class III & -- & Class I/II & Transitional Disk & Class II\\
\addtocounter{footnote}{-2}
Spectral Type\footnote{2} \footnote{3} & K2 & -- & K2 & -- &  & M4 & M3 \\ 
Alt. Designation & \object{HD 288313A}  & -- & HD 288313A  & -- & \object{HD 288313B} & \object{[OH83] L1622 6} & \object{L1622-6N}  \\ \hline
\end{tabular}}}

%% file: table_xspec_fit_v6.tex
%Note: Starting with version 3 of this table, the 515 Aa data is different cause I used a new aperture size of 0.25 to match that of Ab.
%version 4 is getting rid of the LX due to the extraction specific issue and adding int he scaling factor lx (3.4*lx) based on psf area.
%version 5 is getting rid of the core extraction regions from joels advice.  I used excel to do it quick so check for errors
%v6 is converting the flux to 1E-14 units
{\renewcommand{\arraystretch}{1.5}
%\resizebox{\textwidth}{!}{
\begin{tabular}{ c  c  c  c }

%\begin{tabular}{| c | c  c | c |}
\hline

&HBC 515 Aa Wing&HBC 515 Ab Wing&HBC 515 C  \\ \hline
$N_{H}$ [$\times10^{21}$ cm$^{-2}$]&2.51$_{-0.52}^{+0.71}$&2.82$_{-0.72}^{+1.09}$&2.05$_{-1.15}^{+1.68}$ \\
$kT_{1}$ [keV]&0.432$_{-0.066}^{+0.065}$&0.328$_{-0.055}^{+0.079}$&0.381$_{-0.072}^{+0.13}$  \\
$kT_{2}$ [keV]&2.37$_{-0.363}^{+0.565}$&1.92$_{-0.236}^{+0.261}$&2.14$_{-0.719}^{+2.031}$  \\
$ Norm._{1}$  [$\times10^{-4}$  cm$^{-5}$]&3.56$_{-1.83}^{+3.4}$&1.94$_{-1.0}^{+2.42}$&1.53$_{-1.05}^{+2.63}$  \\
$ Norm._{2}$ [$\times10^{-4}$ cm$^{-5}$]&4.01$_{-0.66}^{+0.56}$&3.87$_{-0.50}^{+0.52}$&0.67$_{-0.26}^{+0.29}$  \\
\footnotemark[1]Ne&1.01$_{-0.4}^{+0.71}$&1.48$_{-0.57}^{+1.04}$&0.722$_{-0.35}^{+0.98}$  \\
\footnotemark[1]Fe&0.14$_{-0.07}^{+0.15}$&0.31$_{-0.14}^{+0.19}$&0.06$_{-0.05}^{+0.13}$  \\
\footnotemark[2]Obs. $F_{X}$ [$\times$ 10$^{-14}$ erg s$^{-1}$cm$^{-2}$]&43.5$_{-4.2}^{+0.90}$&33.7$_{-3.5}^{+0.90}$&9.29$_{-2.41}^{+0.63}$  \\
Intrin. $L_{X}$ [erg s$^{-1}$]&\footnotemark[3]6.5$\times 10^{31}$&\footnotemark[1]5.4$\times 10^{31}$&3.75$\times 10^{30}$  \\
Red. $\chi^{2}$ &	1.00	&0.87	& 0.86 \\ \hline

\end{tabular}}

%% file: table_SMA_hbc515_v3.tex
%v2 just reduced the sig figs
%added the RA and DEC of cont detection for ref

{\renewcommand{\arraystretch}{1.5}

%\resizebox{\textwidth}{!}{
\begin{tabular}{ c  c  c  c  c  c }
%\begin{tabular}{| c | c | c | c | c | c |}
\hline

& HBC 515Aa & HBC 515Ab & HBC 515 B & HBC 515 C & HBC 515D\footnotemark[1] \\ \hline
Continuum Centroid RA & -- & -- & 05:54:03.27 & 05:54:01.90 & --  \\
Continuum Centroid DEC & -- & -- & +01:40:26.49 &+01:40:27.38  & -- \\
1.3 mm Continuum Flux [mJy] & $<$ 2.9 & $<$ 2.9 & 29.9 & 6.5 & $<$ 7.7\\
$^{12}$CO Flux [mJy km s$^-1$] & $<$ 1.2 & $<$ 1.2 & $<$ 1.2 & $<$ 1.2 & $<$ 3.2 \\ 
$^{13}$CO Flux [mJy km s$^-1$] & $<$ 1.2 & $<$ 1.2 & $<$ 1.2 & $<$ 1.2 & $<$ 3.2\\
Dust mass [M$_E$] & $<$ 13.4 & $<$ 13.4 & 139.5 & 30.4 & $<$ 36.0 \\
Total Disk Mass [M$_J$] & $<$ 4.2 & $<$ 4.2 & 44.3 & 9.6 & $<$ 11.4 \\ \hline

\end{tabular}}